\documentclass[prl,showpacs,superscriptaddress,twocolumn,floatfix]{revtex4-1}
\usepackage[pdftex]{graphicx}  
\usepackage{graphicx, xcolor}
\usepackage{dcolumn}
\usepackage{bm}
\usepackage{amsmath,amsthm,amssymb}
\usepackage{mathtools}
\usepackage{color}
\usepackage{verbatim}

\definecolor{darkGreen}{RGB}{0,110,0}
\definecolor{darkBlue}{RGB}{0,0,130}
\usepackage[colorlinks,citecolor=darkGreen,linkcolor=darkBlue,urlcolor=blue,hyperindex]{hyperref}

\newcommand{\beginsupplement}{%
        \setcounter{table}{0}
        \renewcommand{\thetable}{S\arabic{table}}%
        \setcounter{figure}{0}
        \renewcommand{\thefigure}{S\arabic{figure}}%
     }

\newcommand{\ts}{\textsuperscript}

\begin{document}
\title{Energy transport in a disordered spin chain with broken U(1) symmetry:\ Diffusion, subdiffusion, and many-body localization}

\author{M. Schulz$^{*}$}
\affiliation{SUPA, School of Physics and Astronomy, University of St Andrews, North Haugh, St Andrews, Fife KY16 9SS, United Kingdom}
\affiliation{Max Planck Institute for the Physics of Complex Systems, N{\"o}thnitzer Str.\ 38, 01187 Dresden, Germany}
\author{S.R. Taylor$^{*}$}
\affiliation{SUPA, School of Physics and Astronomy, University of St Andrews, North Haugh, St Andrews, Fife KY16 9SS, United Kingdom}
\author{C.A. Hooley}
\affiliation{SUPA, School of Physics and Astronomy, University of St Andrews, North Haugh, St Andrews, Fife KY16 9SS, United Kingdom}
\author{A. Scardicchio}
\affiliation{The Abdus Salam International Center for Theoretical Physics, Strada Costiera 11, 34151 Trieste, Italy}
\affiliation{INFN Sezione di Trieste, Via Valerio 2, 34127 Trieste, Italy}
\date{Wednesday 23rd May 2018}


\begin{abstract}
We explore the physics of the disordered XYZ spin chain using two complementary numerical techniques:\ exact diagonalization (ED) on chains of up to 17 spins, and time-evolving block decimation (TEBD) on chains of up to 400 spins.  Our principal findings are as follows.  First, we verify that the clean XYZ spin chain shows ballistic energy transport for all parameter values that we investigated.  Second, for weak disorder there is a stable diffusive region that persists up to a critical disorder strength that depends on the XY anisotropy.  Third, for disorder strengths above this critical value energy transport becomes increasingly subdiffusive.  Fourth, the many-body localization transition moves to significantly higher disorder strengths as the XY anisotropy is increased.  We discuss these results, and their relation to our current physical picture of subdiffusion in the approach to many-body localization.
\end{abstract} 

\maketitle

\paragraph{Introduction.}
Although quantum mechanics is over a hundred years old, some of its most striking predictions about macroscopic systems have been overlooked until recently.  Now, however, technological progress in isolating and controlling nano- and mesoscopic quantum systems \cite{Will2013,Sattler2016} has led to renewed interest in their fundamental properties.  These newly available experimental avenues, and the associated computational and analytical progress, are once again bringing questions about the quantum mechanics of macroscopic systems to the fore.

One recent prediction is the typicality of a localized phase in strongly disordered quantum systems, an effect known as many-body localization (MBL) \cite{Basko:2006hh,oganesyan2007localization,pal2010many,nandkishore2015many,abanin2017recent}. For non-interacting systems, it has long been known from the work of Anderson \cite{Anderson1958} (and the large amount of numerical and analytical work that followed \cite{evers2008anderson}) that, when the disorder is sufficiently strong, transport stops. Examples include impurity-band electrons in a semiconductor at sufficiently low densities \cite{mott1961theory} and waves propagating in a medium with an irregular dielectric constant \cite{akkermans2007mesoscopic}.

Recent works aimed at determining what localization means for interacting systems have shown the phenomenology of these `transportless' MBL systems to be quite rich.  Their novel physics includes the slow but continued growth of entanglement measures due to dephasing \cite{vznidarivc2008many,bardarson2012unbounded,Serbyn:2013he,John2015TotalCorrelations,iemini2016signatures} and the emergence of integrability \cite{serbyn2013local,huse2013phenomenology,ros2015integrals,imbrie2017review}.  The latter has important implications for future technologies, in particular for quantum computation \cite{altshuler2010anderson,Laumann2015,PhysRevLett.118.127201}.

While there are ongoing debates about the differences between one-dimensional chains and higher-dimensional lattices \cite{chandran2016many,de2017stability,huveneers2017classical}, and about the existence or non-existence of a transition in energy at fixed disorder strength \cite{luitz2015many,laumann2014many,de2016absence}, the basic physics of the MBL phase is nonetheless fairly well understood by now.  By contrast, very little progress has been made on the properties of the transition between the ergodic and MBL phases, and in particular the region immediately preceding it on the low-disorder side.  Numerics in the critical and pre-critical regions of the isolated system scarcely converge, and the critical exponents that emerge from a scaling analysis appear to be ruled out by general considerations \cite{chandran2015finite,luitz2015many,pietracaprina2017entanglement}.

It is thus useful to observe that one can access much larger system sizes by considering open-system dynamics. Previous papers have pursued this idea to characterize transport in XXZ spin chains, where the $z$-projection of the total spin is conserved:\ spin transport in \cite{Znidaric2016Diffusive,Reichman2014Absence,AgarwalAnomalousDiffusion}, and energy transport in \cite{Vipin2015} though with severely limited numerics. To summarize the results of \cite{Znidaric2016Diffusive}, there is a small region of diffusive transport and a large, pre-critical region of subdiffusive transport.

\begin{figure}
	\center{\includegraphics[width=1.0\linewidth]{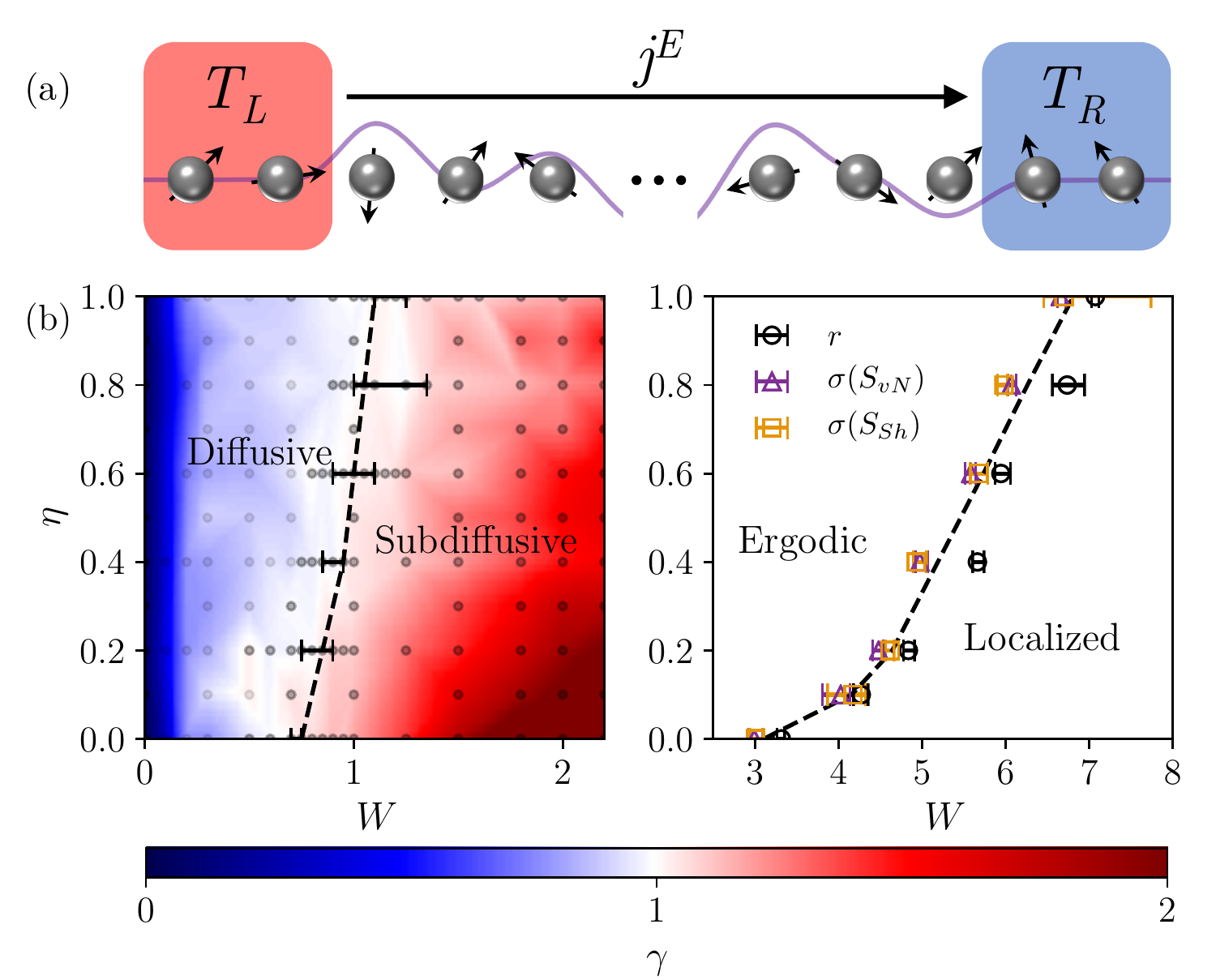}}
	\caption{(a) A disordered spin-1/2 XYZ chain, with Lindblad driving applied to the pair of spins at each end to impose a temperature gradient.  In our time-evolving block decimation (TEBD) studies, we time-evolve such a system until it reaches its non-equilibrium steady state (NESS).  We supplement that analysis by exact diagonalization (ED) studies on closed (and much shorter) chains.  (b) The resulting `phase diagram' of the disordered spin-1/2 XYZ chain, for an Ising anisotropy of $\Delta = 1.2$.  Here $\eta$ is the XY anisotropy of the exchange interaction between the spins, and $W$ is the strength of the random-field disorder.  In the left-hand panel, the dashed line shows the border between diffusive and subdiffusive energy transport determined from our TEBD studies, and the bars are error estimates. The color scale shows the transport exponent $\gamma$ estimated via interpolation between the numerically determined values, which are indicated by gray points. The right-hand panel shows the location of the MBL transition, determined by three different analyses of our ED results:\ the crossover in level statistics from random-matrix to Poissonian, $r$; the peak in the standard deviation of the Shannon entropy, $\sigma(S_{\rm Sh})$; and the peak in the standard deviation of the von Neumann entropy, $\sigma(S_{\rm vN})$.}
	\label{Fig:Phase_Diagram}
\end{figure}

In this paper we investigate the physics of the disordered spin-$1/2$ XYZ chain.  We choose this model because it is a quantum spin chain in which all conservation laws are violated except energy.  In particular, the U(1) symmetry of the XXZ model, which corresponds in a fermionic picture to fermion number conservation, is broken in the XYZ model.  Our investigation employs two complementary techniques.  First, we time-evolve open chains of up to 400 spins using time-evolving block decimation (TEBD).  This method, shown schematically in Fig.~\ref{Fig:Phase_Diagram}(a), gives us access to the transport properties of the system at weak-to-intermediate disorder strengths, including the subdiffusive region.  Second, we use exact diagonalization (ED) on chains of up to 17 spins, which gives us access to the spectral properties of the system at strong disorder, including the MBL transition itself.

\paragraph{Summary of main results.} Our principal findings are as follows:

First, we verify that in the absence of disorder ($W=0$) the transport is ballistic \cite{Zotos1997transport}, in contrast with the classical model \cite{PhysRevB.72.140402} where the non-linear interaction between the spin modes causes spin waves to diffuse. We attribute this behavior to the integrability of the quantum model \cite{Baxter1982}, as non-integrable or classical spin chains typically show diffusive transport (see for example \cite{mukerjee2006statistical,oganesyan2009energy,dymarsky2018bound}).
 
Second, for weak but non-zero disorder ($0 < W \lesssim 0.7$) there is a region in which energy transport is diffusive.  This diffusive region persists up to a finite critical disorder strength, $W_{c1}(\eta)$, which depends on the XY anisotropy $\eta$ (i.e.\ on how strongly the U(1) symmetry of the XXZ chain is broken).

Third, for increasing disorder strengths $W > W_{c1}(\eta)$ energy transport becomes increasingly subdiffusive, while increasing the XY anisotropy $\eta$ counteracts this effect and brings the system back towards the regime of diffusive energy transport.  We can follow this behavior up to disorder strengths of $W \approx 2.2$, where we see subdiffusive exponents up to $\gamma \approx 2.7$.

Fourth, the system exhibits an MBL transition at a disorder strength $W_{c2}(\eta)$, which increases significantly as the XY anisotropy $\eta$ is increased.  Due to the abovementioned lack of a U(1) symmetry in the XYZ chain, this transition cannot be thought of as directly following from the arguments for localization of \cite{Basko:2006hh}.  It is, however, in line with the most recent research on the topic which relies less on the particle interpretation \cite{imbrie2014many} and more on non-proliferation of resonances.
We determine $W_{c2}$ via ED analysis of chains with lengths up to $L=17$ spins, using the standard tests of the eigenstates and spectrum of the Hamiltonian \cite{oganesyan2007localization, luitz2015many, Kjall2014, Torres-Herrera2017}.

A phase diagram summarizing these results is shown in Fig.~\ref{Fig:Phase_Diagram}(b).  In the remainder of this paper we present the details of the model under study and the methods we use, and then proceed to discuss each of these results in turn.

\paragraph{Model.}
The Hamiltonian of the disordered XYZ spin chain is
\begin{eqnarray}
H & = & \sum_{n=1}^{L-1} \Big[  (1+\eta)s_n^x s_{n+1}^x + (1-\eta) s_n^y s_{n+1}^y + \Delta s_n^z s_{n+1}^z \Big] \nonumber \\
& & +\, \sum_{n=1}^L h_n s_n^z.
\label{eq:H_XYZ}
\end{eqnarray}
Here $s^{\alpha}_{n} = \frac{1}{2}\sigma^{\alpha}_{n}$ are spin-1/2 operators ($\sigma^{\alpha}_{n}$ are Pauli matrices), $\eta$ is the XY anisotropy of the coupling (the parameter that breaks the U(1) symmetry of the XXZ model), $\Delta$ is the Ising anisotropy, and $h_n \in \left[-W,W\right]$ are uncorrelated disorder fields randomly drawn from a uniform distribution. The $\eta \to 0$ limit of this model is the well-studied XXZ spin chain; $\eta \neq 0$ introduces a term equal to $\eta \sum_n \left( s^+_n s^+_{n+1} + s^-_n s^-_{n+1} \right)/2$, which violates the conservation of the $z$-component of the total magnetization.  In the fermion language this corresponds to a nearest-neighbor pairing term.

\paragraph{Methods.} We use two complementary methods:\ TEBD on open chains, and ED on closed ones.  In our TEBD studies we couple the ends of the chain to two thermal baths at different temperatures, and describe the time-evolution of the resulting open system using the Lindblad equation \cite{Breuer2002}
\begin{equation}
\frac{d\rho}{dt}\ = -i\left[H,\rho\right] +\kappa \left\{ \vphantom{c^\dagger} \mathcal{L}_L(\rho) +  \mathcal{L}_R(\rho)\right\}. \label{lindblad}
\end{equation}
The first term on the right-hand side of (\ref{lindblad}) describes the coherent dynamics; the Lindblad terms $\mathcal{L}_{L}(\rho)$  and $\mathcal{L}_{R}(\rho)$ correspond to the left and right reservoirs respectively, and $\kappa$ is the strength with which we couple them to the chain.  We apply a two-site thermal driving protocol that has been used in similar transport studies \cite{Prosen2009,Znidaric2010}, which drives an isolated pair of spins to a thermal state with temperature $T$, $\rho \propto \exp(- H / T)$. We drive the pair of spins on the left-hand end of the chain towards a high temperature $T_L$, and the right-hand pair towards a lower temperature $T_R$, as depicted in Fig.~\ref{Fig:Phase_Diagram}(a). For the remainder of this paper we use the target temperatures $T_L = \infty$ and $T_R = 20$.

\begin{figure}
\center{\includegraphics[width=1.0\linewidth]{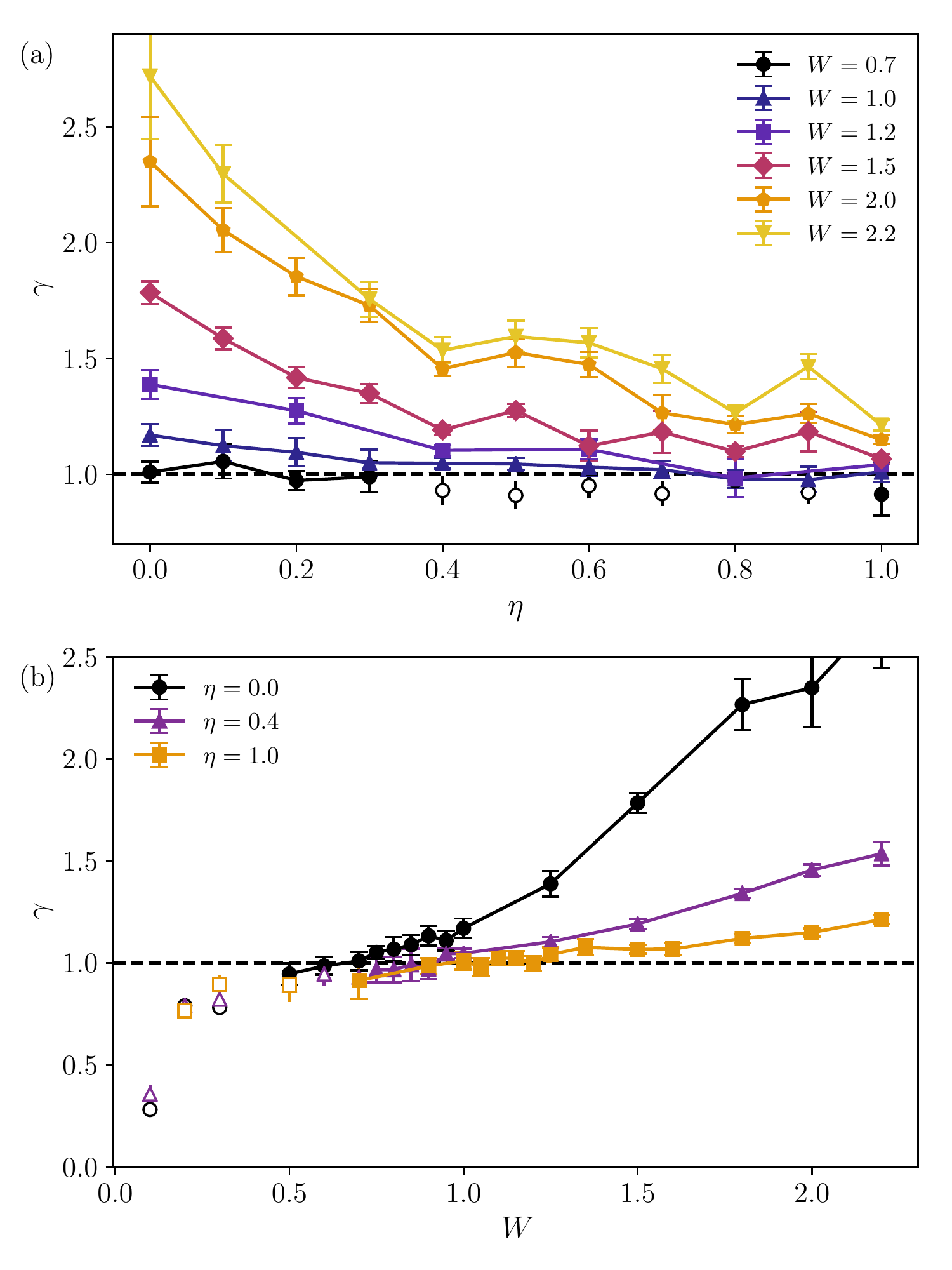}}
\caption{The energy-transport exponent $\gamma$ in the disordered spin-1/2 XYZ chain at weak to moderate disorder strengths, as determined from our TEBD numerical results.  All results reported are for an Ising anisotropy of $\Delta=1.2$.  (a) The exponent $\gamma$ as a function of the XY anisotropy $\eta$, for various values of the disorder strength $W$.  $\gamma = 1$ corresponds to diffusive energy transport; for $\gamma > 1$, energy transport is subdiffusive.  (b) The exponent $\gamma$ as a function of the disorder strength $W$, for various values of the XY anisotropy $\eta$.  The open symbols in both panels indicate cases in which the chain was not long enough to achieve fully diffusive behavior, and these points should therefore be disregarded (see Fig.~\ref{Fig:Scaling} and corresponding text).}
\label{Fig:Disordered_XYZ}
\end{figure}

We then solve (\ref{lindblad}) via TEBD to find the nonequilibrium steady state (NESS) energy current in the chain, $j^E$, for various values of its length, $L$.  The TEBD approach permits us to reach very large system sizes of up to $L = 400$ spins, avoiding the severe finite-size effects described in \cite{Znidaric2016Diffusive}. Details of our simulation can be found in the Supplemental Material.
We then analyze the scaling of $j^E(L)$ with the length of the system $L$. In the delocalized region preceding the MBL transition we expect the current to scale as $j^E \sim L^{-\gamma}$, where $\gamma=0$ corresponds to ballistic transport, $\gamma=1$ to diffusion, and $\gamma > 1$ to anomalous subdiffusive transport \cite{Znidaric2016Diffusive}.  The results of this analysis are shown in Fig.~\ref{Fig:Disordered_XYZ}.

\begin{figure}
\center{\includegraphics[width=1.0\linewidth]{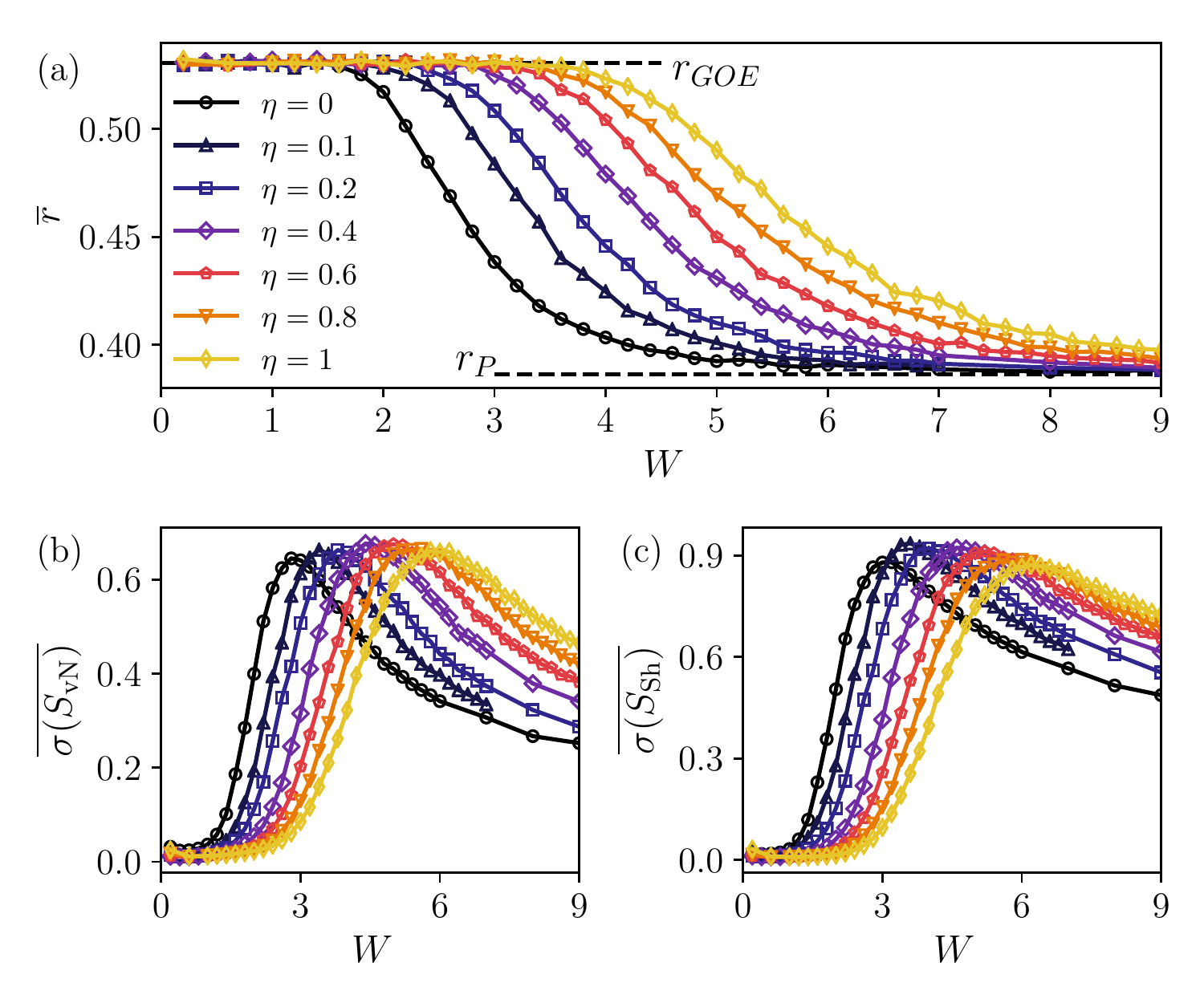}}
\caption{Locating the MBL transition via exact diagonalization.  This figure shows the (a) level statistics $r$ parameter, (b) the standard deviation of the entanglement entropy distribution, and (c) the standard deviation of the Shannon entropy distribution, all as a function of disorder strength $W$ for several values of the XY anisotropy parameter $\eta$.  These results were obtained from exact diagonalization of the Hamiltonian for a disordered XYZ spin chain of length $L=15$ with periodic boundary conditions.  The error bars are smaller than the symbol size.}
\label{Fig:ED}
\end{figure}

Because the convergence of our TEBD method worsens at stronger disorder, we cannot use it all the way to the MBL transition.  Therefore, we also perform ED studies on short, closed chains (up to $L=17$ spins for the XXZ model and $L=16$ spins for the XYZ model) with periodic boundary conditions.
We identify the location of the MBL transition using the crossover from random-matrix to Poissonian statistics in the eigenenergy spectrum and the peaks in the fluctuations of the Shannon entropy $S_{\mathrm{Sh}}$ and the half-chain entanglement entropy $S_{\mathrm{vN}}$.
We evaluate these quantities using the 200 eigenstates closest to the middle of the many-body energy spectrum, and then average over disorder realizations. We then determine the location of the MBL transition by performing a finite-size scaling analysis of the disorder-averaged results.  The results of this analysis are shown in Fig.~\ref{Fig:ED}.
\begin{figure}
\center{\includegraphics[width=1.0\linewidth]{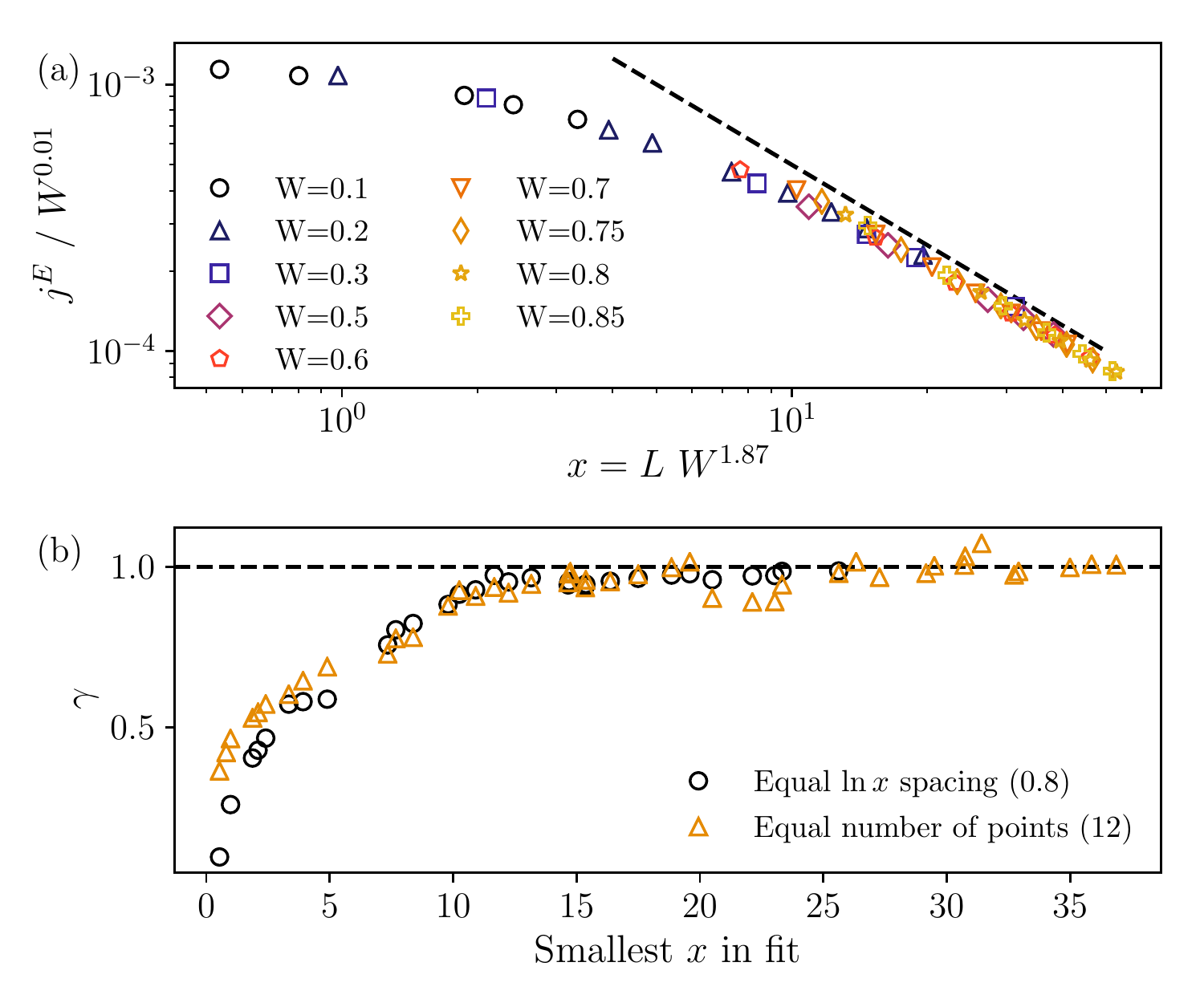}}
\caption{Testing whether our chains are long enough for the scaling limit to have been reached.  In this example the XY anisotropy parameter $\eta = 0.4$.  (a) The numerical collapse of the energy current $j^E$ as a function of chain length $L$ onto a single curve under suitable scaling by the disorder strength $W$.  Note the typical crossover from ballistic behavior in short chains to diffusive behavior --- indicated by the dashed line --- in longer ones. (b) The `running exponent' $\gamma(x)$, determined from tangential power-law fits to the universal curve, showing that $\gamma$ reaches the diffusive value of 1 above the critical length scale $x^\star \approx 25-30$.}
\label{Fig:Scaling}
\end{figure}

\paragraph{No disorder:\ Ballistic energy transport.}
In the limit of no disorder ($W=0$), we find that the energy current $j^E$ is independent of the length of the system, which signals that the energy transport is ballistic; this is consistent with previous work on the XYZ model \cite{Zotos1997transport}. Ballistic energy transport has been linked to the integrability of quantum systems \cite{Mendoza2015}, a characteristic which is also visible in the Poissonian statistics of the Hamiltonian's eigenenergy spectrum \cite{Caux2011}. For $0 < \Delta \leq 2$ and $0 < \eta \leq 1$ we find that the average of $r$ falls close to the Poissonian value $r_P = \ln4 -1$ over the entire spectrum. Details of this analysis can be found in the Supplemental Material.

\paragraph{Weak disorder:\ Stable diffusive phase.}
At weak but non-zero disorder, $0< W < W_{c1}(\eta)$, the transport is diffusive. This has previously been shown for spin and energy transport in the XXZ chain ($\eta = 0$) \cite{Mendoza-Arenas2018}, and we report it here in the XYZ case. The diffusive phase is not materially altered when the XY anisotropy is increased, except insofar as it extends to stronger disorder, i.e.\ $W_{c1}(\eta)$ increases with $\eta$ (see Fig.~\ref{Fig:Disordered_XYZ}(b)). We explain in the Supplemental Material how we obtain $W_{c1}(\eta)$.

As in previous studies, we find severe finite-size effects in the results at weak disorder, with the asymptotic scaling behavior of $j^E(L)$ observed only for values of $L$ exceeding a critical length $L^\star$, where $L^\star$ increases with decreasing $W$. For a ballistic-to-diffusive crossover, it has been shown that this length scale should scale as $L^\star \sim W^{-2}$ \cite{Znidaric2016Diffusive}.  If we apply our analysis na{\"\i}vely to a chain of length $L < L^\star$, it yields an exponent $\gamma < 1$, and thus falsely suggests superdiffusive energy transport.

However, we can use the scaling properties of $j^E(L)$ to test whether the scaling regime has been reached in any given case.  In Fig.~\ref{Fig:Scaling}(a) we demonstrate that, by scaling the data using $x \equiv LW^\nu$ and $y \equiv j^E W^{\delta-\nu}$, it is indeed possible to collapse all points onto a single universal curve.  For the example shown, $\eta = 0.4$, the best empirical scaling exponent is $\nu \approx 1.87$, in reasonable agreement with the predicted value of 2. We also find that $\nu - \delta = 0.01$, which is close to the predicted behavior of $\nu = \delta$ \cite{Znidaric2016Diffusive}.

On the basis of this analysis, we indicate via open symbols in Fig.~\ref{Fig:Disordered_XYZ} those cases where the scaling regime has not been reached, and where we are therefore confident that the reported value of $\gamma$ is not reflective of the thermodynamic limit.  Further details of how we identify these points may be found in the Supplemental Material.

\paragraph{Intermediate disorder:\ Subdiffusive energy transport.}
We find that the disordered XYZ model exhibits subdiffusive energy transport at $W > W_{c1}(\eta)$. In contrast to the diffusive region, in the subdiffusive phase the transport exponent $\gamma$ varies continuously as a function of both the disorder strength $W$ and the XY anisotropy $\eta$.  This variation shows two main trends.  First, as shown in Fig.~\ref{Fig:Disordered_XYZ}(a), a larger $\eta$ results in a smaller $\gamma$, i.e.\ breaking the U(1) symmetry pushes the system back towards diffusive transport.  Second, as shown in Fig.~\ref{Fig:Disordered_XYZ}(b), increasing disorder strength $W$ leads to an increased value of $\gamma$ for all values of $\eta$, i.e.\ increasing disorder pushes the system further away from the diffusive regime.  While we cannot follow this behavior all the way to the MBL transition, the location of which we determine by other means, we expect that $\gamma$ would diverge there.

\paragraph{Strong disorder:\ Many-body localization.}
The disorder-averaged level statistics parameter $r$, and the standard deviations of two types of entropy fluctuation $\sigma(S_{\mathrm{vN}})$ and $\sigma(S_{\mathrm{Sh}})$, are shown as a function of disorder strength in Fig.~\ref{Fig:ED}.  All three measures demonstrate a pronounced increase of the critical disorder strength for the MBL transition, $W_{c2}$, as the XY anisotropy parameter $\eta$ is increased.
The phase diagram in Fig.~\ref{Fig:Phase_Diagram}(b) shows the approximate position of the MBL transition according to a scaling analysis of these data. The scaling analysis was performed by numerically collapsing the data for different chain lengths to a function of the form $g ( L^{1/\nu} [ W - W_{c2} ]  )$, where $\nu$ and $W_{c2}$ are fitting parameters, as in previous work \cite{luitz2015many}. We note that, as found in similar studies, the exponent $\nu < 2$, contrary to established predictions \cite{chandran2015finite}.

\paragraph{Discussion.}  In this paper, we have provided evidence that there are four phases in the disordered spin-$1/2$ XYZ chain:\ a ballistic phase at zero disorder; a diffusive phase for a finite range of disorder from $0^{+}$ to a critical value $W_{c1}(\eta)$; a subdiffusive phase for a finite range of disorder from $W_{c1}(\eta)$ to the MBL transition $W_{c2}(\eta)$; and a many-body localized phase for disorders above $W_{c2}(\eta)$.  Importantly, the model that we have studied takes us beyond cases --- such as the previously studied XXZ chain --- that can be thought of in terms of the strongly-interacting dynamics of a fixed number of particles. The XYZ model breaks the U(1) symmetry in a controlled way, and this allows us to observe the changing behavior of the system as we interpolate from the XXZ chain to other models (such as the transverse-field Ising model) which exist in separate regions of parameter space.

The essential physics can be summed up in two short phrases:\ Disorder tends to localize; XY anisotropy tends to delocalize.  How should we understand the latter effect?  One way is to think in the fermionic picture, in which the XY anisotropy $\eta$ appears as a pair-creation (and of course a partner pair-annihilation) term.  This means that the system, in its time-evolution, can visit sectors with other fermion numbers, which it could not in the XXZ case.  Barring significant phase-coherence effects between the states in the $N$ and $N+2$-particle sectors (which there seems to be no reason to expect), this opens up different channels for energy transport, and thus would be expected to enhance the delocalization of energy density excitations.  We plan to present further details of this argument in a forthcoming publication \cite{msca}.

\paragraph{Acknowledgments.} We would like to thank Marko \v{Z}nidari\v{c}, Jonathan Keeling, Roderich Moessner, and Heiko Burau for helpful discussions/comments throughout this project. MS and SRT acknowledge financial support from the CM-CDT under EPSRC (UK) grants EP/G03673X/1 and EP/L015110/1.  CAH acknowledges financial support from the TOPNES programme under EPSRC (UK) grant number EP/I031014/1.  This research was supported in part by the National Science Foundation under Grant No.\ NSF PHY-1125915.  Part of this work was performed at the Aspen Center for Physics, which is supported by National Science Foundation grant PHY-1066293.  AS's research is in part supported by a Google Faculty Award.

\bibliography{MBLbib}

\appendix
\section{Supplemental Material}

\beginsupplement

\subsection{The clean XYZ model}
\label{supp:clean}

In the zero-disorder limit the XYZ model has been shown to be solvable by Bethe Ansatz methods. However, the ``reference state'' is not known, so only a limited number of exact results are available \cite{Baxter1982,Inami1994}. Integrability is linked with ballistic energy transport as well as Poissonian level statistics. However, as far as we are aware, there have hitherto been no transport or eigenvalue studies of this sort for the XYZ model.

In the main text we stated that the clean XYZ model shows ballistic transport and Poissonian level statistics over the entire parameter space of $\Delta$ and $\eta$ that was studied in this paper. In the following we provide evidence for this claim as well as giving details on the numerical methods.

\begin{figure}
	\center{\includegraphics[width=1.0\linewidth]{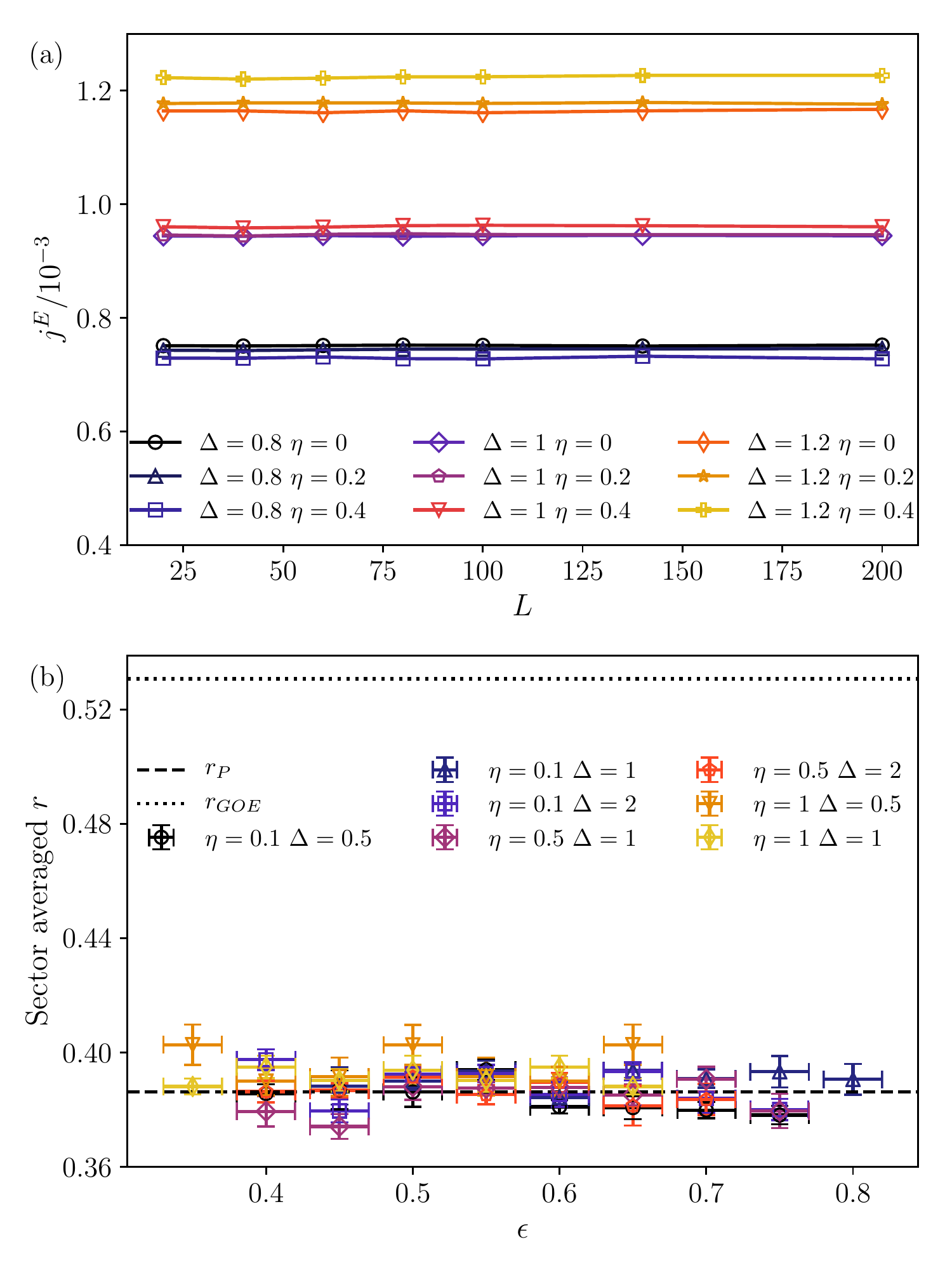}}
	\caption{Properties of the XYZ model without disorder. (a) TEBD results for the dependence of the energy current $j^E$ on the chain length $L$, showing ballistic
		transport for several values of the Ising anisotropy $\Delta$ and the XY anisotropy $\eta$. (b) ED results for the energy-resolved average gap-ratio parameter $r$ for several values of $\Delta$ and 
		$\eta$, averaged over symmetry sectors, showing that the spectrum follows Poissonian statistics.}
	\label{Fig:Clean}
\end{figure}

Fig.~\ref{Fig:Clean}(a) shows our TEBD results for the energy current $j^E$ as a function of the chain length $L$, for a range of $\eta$ and $\Delta$.  Each curve clearly shows that $j^E$ is independent of the chain-length $L$, as expected for ballistic transport. Our results are therefore consistent with the established analytical predictions for spin chains in the absence of disorder \cite{Zotos1997transport}.

Fig.~\ref{Fig:Clean}(b) shows our ED results for the gap-ratio parameter in a chain of length $L=16$ with open boundary conditions, for a range of $\eta$ and $\Delta$.  The results are averaged over small regions of the spectrum ($\pm 0.02$ of the bandwidth around a target $\epsilon$), all of which fall close to the Poissonian value which would be expected of an integrable model. Horizontal bars show the region of the spectrum over which the average is taken, and vertical bars show the standard error of the average between the different symmetry sectors of the Hamiltonian.  (These different sectors are distinguished by (a) the parity of the number of up spins, (b) their parity under reflection across the centre of the chain, and (c) their parity under inversion of all spins.)  Only results where the interval contains at least 400 levels are shown.

\subsection{TEBD:\ details of method}

\begin{figure*}
	\center{\includegraphics[width=\textwidth]{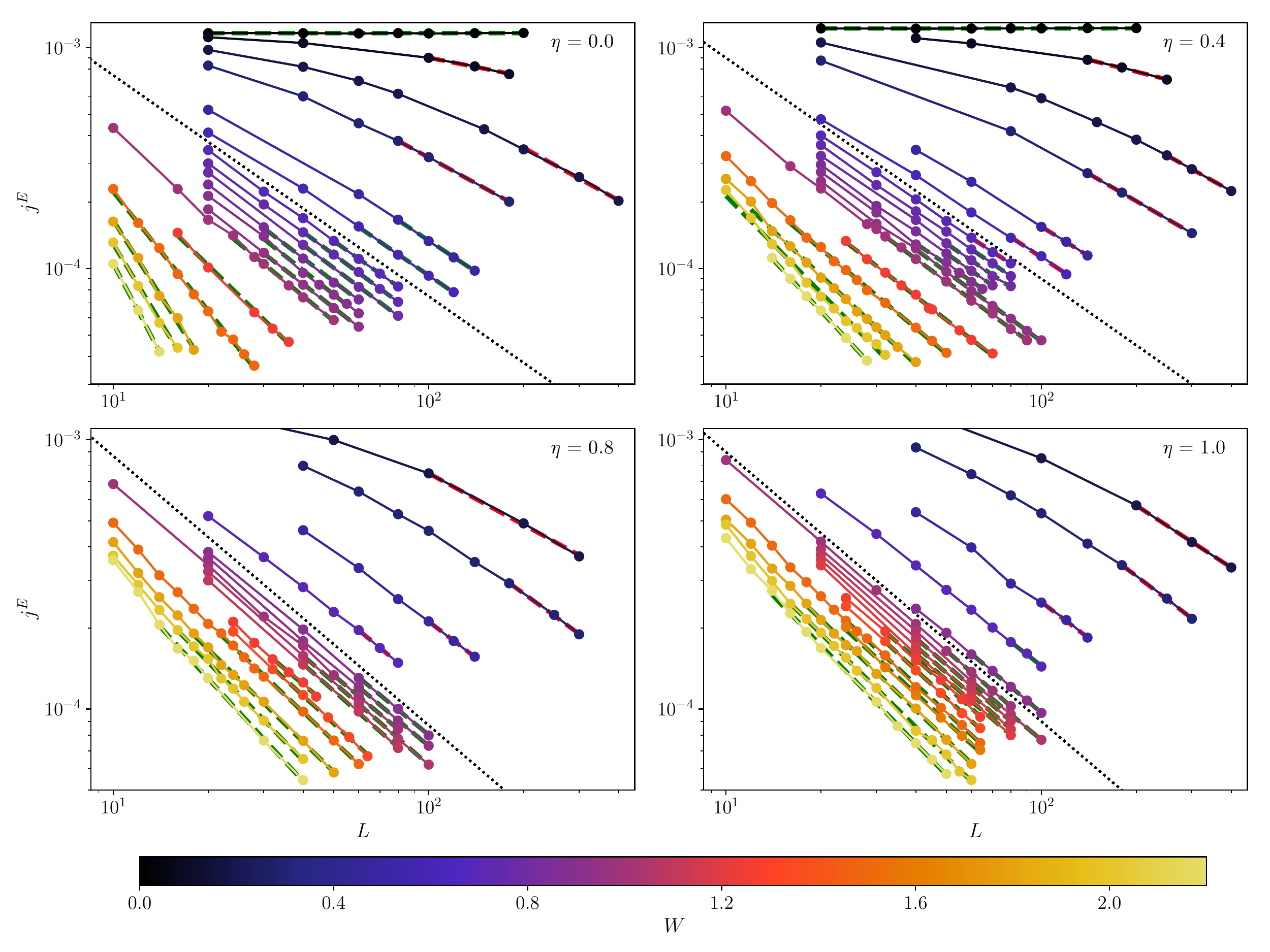}}
	\caption{Transport properties of the XYZ model for a range of disorder strengths $W$, including zero. As a complement to Fig.~\ref{Fig:Disordered_XYZ} in the main text, we show here the underlying current scaling for different values of the XY anisotropy $\eta$. Each panel shows the NESS disorder-averaged energy current $j^E$ as a function of chain length $L$. The finite-size effects discussed in the main text are clearly visible for small $W$. The colored dashed lines indicate the power-law fit we use to obtain the transport exponent $\gamma$. The length of the dashed line shows the range of system sizes that were used:\ either all chains longer than the estimated $L^{\star}$ (green), or the last three obtained data points (red). The latter are indicated by hollow symbols in the main part of the paper. The dotted line corresponds to a transport exponent of $\gamma = 1$.}
	\label{Fig:curr_scaling}
\end{figure*}

To study the energy transport across the disordered chain, we simulate the non-equilibrium configuration depicted in Fig.~\ref{Fig:Phase_Diagram}(a). We use two-site bath operators that are designed to induce a thermal state of a given target temperature $T$ on a pair of isolated spins. The numerical details of how this can be achieved with Lindblad terms can be found in refs.\  \cite{Prosen2009,Znidaric2010,Mendoza2015}. The thermal baths are coupled with strength $\kappa$ (in our simulations always $\kappa = 0.5$) to the end spins, where we set $h_i = 0$ ($i \in \{0,1, L-1,L\}$). These end spins are then coupled to the rest of the chain which evolves coherently according to the Hamiltonian $H$ (see \eqref{eq:H_XYZ}).
In order to obtain results for sufficiently long chains, we use a TEBD method to drive the chain until a NESS is obtained. Details of these now widely employed methods can found in refs.\ \cite{Zwolak2004,Verstraete2004,Daley2004,Schollwock2011}. In particular, we encode the density matrix of the system as a state vector and make use of the superoperator formalism to evolve it in time \cite{Zwolak2004,Joshi2013}.

To obtain the NESS for each set of parameters $W$, $\Delta$, and $\eta$, we consider $M$ realizations of the disordered magnetic field $h_n$. For each realization, we take $\rho(0)$ to be a product of completely mixed local density matrices, and we time-evolve it to obtain an approximation to the steady-state density matrix of the lattice, $\rho_{\infty} = \text{lim}_{t\rightarrow \infty} \rho(t)$ \cite{Prosen2009}.  Our TEBD method is a variant of the open source library TeNPyLight \cite{TenPyLight}, which implements the superoperator formalism for states. We use a time-step $dt = 0.4$ and a fourth-order Trotter decomposition \cite{Suzuki1990} for our two-site local updates. At any time, our global density matrix $\rho$ is described by a MPS of matrix dimension of up to $\chi = 300$.

This process is performed until a series of convergence criteria is fulfilled. Spatial homogeneity and temporal uniformity of the energy current are the most suitable indicators of convergence. We choose as our spatial criterion that the standard deviation of the individual currents on every bond $k$ (excluding those subjected directly to the Lindblad driving) relative to the average current be $\sigma(j_k)/\,\overline{j_k} < 2 \%$. The temporal criterion is that the standard deviation of the average current over the previous 100 time-steps is less than $0.3\%$ of the average over the same period.

We then use the obtained NESS as the initial state for a simulation with a higher maximal matrix dimension $\chi$.  As the NESS is unique \cite{Prosen2009}, we make the reasonable assumption that every increase of $\chi$ will bring our numerical approximation of $j_E$ closer to the true value.  We repeat this process of increasing the matrix dimension until the average energy current $\overline{{j}_E}$ is basically independent of $\chi$, i.e.\ when $\Delta \overline{j_E} / \overline{j_E} < 0.4\%$.

If any of the convergence criteria are not satisfied then the result has not converged and the data is excluded from the study. We require that the TEBD must successfully converge to a steady state for at least 98.5\% of disorder realizations, or the data is discarded to avoid biased sampling of the true current distribution.

After the NESS is obtained for $M$ realizations ($M$ ranging $20$ to $600$), the energy current (already averaged along the chain) is averaged over realizations, resulting in a statistical uncertainty of $\sigma(\overline{j_k})/\sqrt{M} \approx 1.5 \%$ or less for most chains ($ < 3-4 \%$ for the strong disorder runs).

\subsection{TEBD:\ current scalings}

In Fig.~\ref{Fig:Phase_Diagram} and Fig.~\ref{Fig:Disordered_XYZ} of the main text, we present values for the energy transport exponent $\gamma$. It is obtained by scaling the NESS energy current as a function of system size $L$ (see Fig.~\ref{Fig:curr_scaling}).
The energy current operator is determined from the continuity equation for the bond energy operator
\begin{eqnarray*}
\tilde{H}_{n,n+1} & = & (1+\eta)s_n^x s_{n+1}^x + (1-\eta) s_n^y s_{n+1}^y + \Delta s_n^z s_{n+1}^z \\
& & \,\,\,\,+\, \frac12 \left( h_n s_n^z + h_{n+1} s_{n+1}^z \right),
\end{eqnarray*}
and in the XYZ model the current operator for the $n$\ts{th} site is:
\begin{eqnarray*}
j^E_n & = & i \left[ \tilde{H}_{n-1,n}, \tilde{H}_{n,n+1} \right] \\
& = & \sum_{\alpha,\beta,\gamma}
J_{\alpha} J_{\gamma} s_{n-1}^{\alpha}
s_n^{\beta} s_{n+1}^{\gamma} \varepsilon_{\alpha\beta\gamma} \\
& & \,\,\,\, +\,\frac12 h_n J_x \left( s_{n-1}^x s_n^y - s_n^y s_{n+1}^x \right)\\
& & \,\,\,\,+ \,\frac12 h_n J_y \left( s_n^x s_{n+1}^y - s_{n-1}^y s_n^x \right)
\end{eqnarray*}
where $\varepsilon_{\alpha\beta\gamma}$ is the Levi-Civita tensor, $\alpha,\beta,\gamma \in \{ x,y,z \}$, and $J_{\alpha}$ is the nearest-neighbor coupling between the $\alpha$ components of the spins.

The diffusion equation for the transport of energy $E$, where $E = \langle \tilde{H}_{n,n+1} \rangle$ corresponds to the bond-energy density as obtained by \eqref{eq:H_XYZ}, reads $j^{E} = - D \nabla E = - D \Delta E/L$. Here $D$ is the diffusion constant, $\nabla E$ is the gradient of $E$, and $\Delta E$ is the difference between its boundary values. In the thermodynamic limit we expect the NESS current scaling to give $j^{E} \sim L^{-\gamma}$. In this case $\gamma = 1$ corresponds to normal diffusive energy transport, i.e.\ Fick's law. When Fick's law breaks down, the transport is no longer diffusive, and we may observe slower subdiffusive ($\gamma > 1$) or faster superdiffusive ($\gamma < 1$) transport.

In Fig.~\ref{Fig:curr_scaling}, the direct application of this analysis is displayed, as well as considerations about finite-size effects (see Fig.~\ref{Fig:Scaling}). From this analysis we further determine a value for the disorder strength $W_{c1}(\eta)$ at which energy transport changes from diffusive to subdiffusive.  We set the upper bound of the transition, i.e.\ the disorder strength at which we are confident that energy transport has become subdiffusive, where our average $\gamma$ is more than two standard deviations away from 1.  The lower bound is set to the disorder value at which the NESS current scaling comfortably falls on top of the universal scaling curve (see Fig.~\ref{Fig:Scaling}). The dashed line in the left-hand panel of Fig.~\ref{Fig:Phase_Diagram}(b) represents the disorder strength at which our average $\gamma$ is more than one standard deviation away from 1.

\subsection{TEBD:\ universal diffusive behavior}

We estimate $x^\star$, the value of $x$ above which the running exponent determined from the scaling of the energy current $j^E(x)$ no longer changes, by performing a tangential fit to the universal diffusive curve formed by the scaled $j^E(L)$ data --- see Fig.~\ref{Fig:Scaling}(a). We perform two fits, the first taken over all of the data points within a region of constant size in $\ln x$ and the second taken over a constant number of data points. The results of these fits are shown in Fig.~\ref{Fig:Scaling}(b) as a function of the smallest $x$ used in each fit.  Our estimate of $x^\star$ is the value at which $\gamma(x)$ has become 1.

This allows us to identify the points in Fig.~\ref{Fig:Disordered_XYZ} where the thermodynamic limit has not been reached, and the value of $\gamma$ is therefore probably underestimated (indicated by hollow markers). We note that the value of $x^{\star}$ increases for increased $\eta$.

\subsection{ED:\ details of method}

\begin{figure*}
	\center{\includegraphics[width=\textwidth]{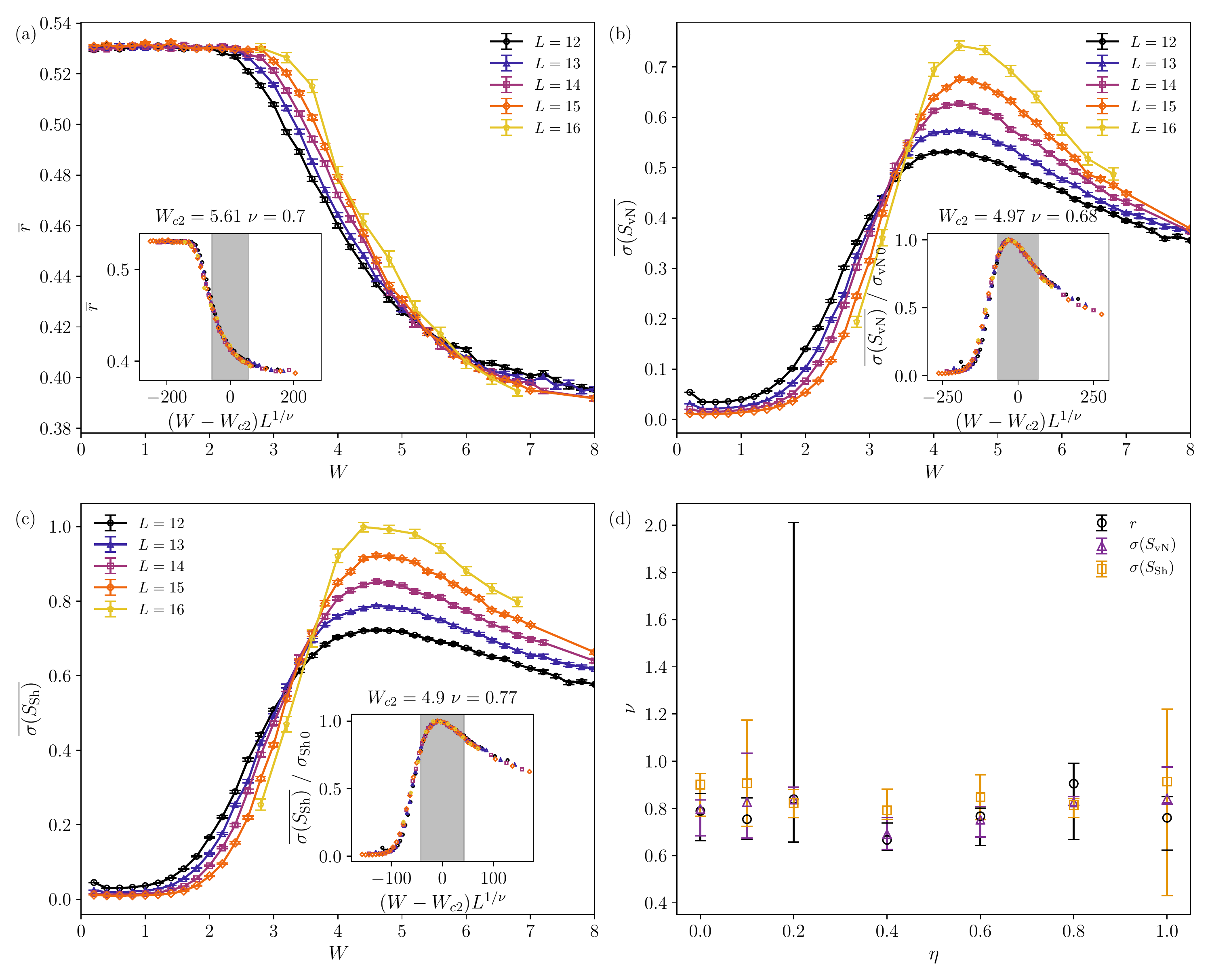}}
	\caption{The scaling analysis of the ED data. The raw data for (a) the gap-ratio parameter $r$, (b) the standard deviation of the entanglement entropy $\sigma(S_{\mathrm{vN}})$, and (c)
			the standard deviation of the Shannon entropy $\sigma(S_{\mathrm{Sh}})$, as a function of disorder strength $W$ for several chain lengths $L$, with XY anisotropy $\eta=0.4$. The insets show
			the numerical collapse performed over a window $W_{c2} \pm 1.7$, indicated by the shaded gray
			region.  (d) The values of the exponent $\nu$ found for each quantity as a function of $\eta$; the points and bars
			indicate the average and range found over all satisfactory fits.}
	\label{Fig:ED_Collapse}
\end{figure*}

We identify the location of the MBL transition using three measures.  First, we locate the crossover from random-matrix to Poissonian statistics in the eigenenergy spectrum, as measured by the gap-ratio parameter $r_n = \min(\delta_n/\delta_{n+1}, \delta_{n+1}/\delta_n)$, where $\delta_n$ is the gap between the $n$\ts{th} and $(n-1)$\ts{th} energy eigenvalues \cite{oganesyan2007localization}.  Second, we locate the peak in the fluctuations of the Shannon entropy $S_{\mathrm{Sh}}= - \sum_i \rho_{ii} \ln \rho_{ii}$, where $\rho$ is the full density matrix of the spin chain.  Third, we locate the peak in the fluctuations of the half-chain entanglement entropy $S_{\mathrm{vN}} = - \mathrm{Tr} (\rho_A \ln \rho_A)$, where $\rho_A$ is the reduced density matrix of the half-chain.

The numerical collapse to a function of the form $g ( L^{1/\nu} [ W - W_{c2} ]  )$ is performed by spline-interpolation of the data onto a rescaled grid, and minimizing the difference between the curves for different $L$. The error bars shown in Fig.~\ref{Fig:Phase_Diagram}(b) correspond to the range of $W_c$ values found by satisfactory collapses performed over a range of windows in $W$ around $W_c$.

Fig.~\ref{Fig:ED_Collapse}(a)-(c) shows the raw data for each length with $\eta=0.4$, and the insets show the collapse of the scaled data. The gray regions show the region in which the minimization was performed.  Fig.~\ref{Fig:ED_Collapse}(d) shows the numerically determined values of $\nu$.  These violate the expected bounds ($\nu \geq 2$ in one dimension), as has also been found in previous studies \cite{luitz2015many, Kjall2014, chandran2015finite}.

\end{document}